\documentclass[prb,twocolumn,preprintnumbers,amsmath,amssymb,floats,citeautoscript,nobalancelastpage,showpacs,superscriptaddress]{revtex4}

\usepackage{graphicx}
\usepackage{dcolumn}
\usepackage{bm}
\usepackage{times}
\usepackage{color}

\begin{document}

\title{Origin of the energy gap in the narrow-gap semiconductor FeSb$_{2}$ revealed by high-pressure magnetotransport measurements}

\author{H.~Takahashi}
\author{R.~Okazaki}
\author{I.~Terasaki}

\affiliation{Department of Physics, Nagoya University, Nagoya 464-8602, Japan}

\author{Y.~Yasui}

\affiliation{Department of Physics, Meiji University, Kawasaki 214-8571, Japan}
\begin{abstract}
To elucidate an origin of the two energy gaps in the narrow-gap semiconductor FeSb$_{2}$, 
we have investigated the effects of hydrostatic pressure on the resistivity, Hall resistance and magnetoresistance at low temperatures. 
The larger energy gap evaluated from the temperature dependence of resistivity above 100 K is enhanced from 30 to 40 meV
with pressure from 0 to 1.8 GPa, as generally observed in conventional semiconductors.
In the low-temperature range where a large Seebeck coefficient was observed, 
we evaluate the smaller energy gap from the magnetotransport tensor using a two-carrier model and 
find that  the smaller gap exhibits a weak pressure dependence in contrast to that of the larger  gap.
To explain the pressure variations of the energy gaps, 
we propose a simple model that the smaller gap is a gap from the impurity level to the conduction band and  
the larger one is a gap between the valence and conduction bands,
suggesting that the observed large Seebeck coefficient is not relevant to electron correlation effects.

\end{abstract}

\pacs{71.55-i, 72.20-i, 72.20.My, 72.20.Pa}

\maketitle

\section{Introduction}

Electron correlations in the Fe-based narrow-gap semiconductors have been extensively studied as a source for 
striking effects to induce unique transport and magnetic behaviors.\cite{kondo1}
In particular, FeSb$_2$ has attracted great interest 
because of the recent observation of a colossal Seebeck coefficient $S\simeq -45$ mV/K 
at around 10 K.\cite{FeSb2}
This compound has two small gaps of $\Delta _{1}\sim 30$ meV and $\Delta _{2}\sim 5$ meV, 
which have been determined from the temperature dependence of resistivity.\cite{FeSb1,FeSb2,FeSb5,FeSb9,FeSb10}
The absolute value of $S$ rapidly increases below 40 K, 
suggesting a possibility that the smaller energy gap $\Delta_{2}$ is formed by an unusual mechanism 
such as a strong electron correlation anticipated in the Kondo insulators,\cite{FeSb2}
in which a steep variation of the density of states near the Fermi level is expected.
The presence of the electron correlation in FeSb$_{2}$ is also suggested 
by several experimental and theoretical studies.\cite{FeSb3,FeSb4,FeSb11}

In the Kondo insulator picture, 
the small energy gaps in FeSb$_{2}$ are thought to arise from the hybridization 
of broad conduction bands of Sb 5$p$ with narrow bands of Fe 3$d$.
In the hybridization gap case,
the value of Seebeck coefficient is expected to be insensitive to amount of impurities.
In highly contrast to the expectation, the reported maximum values of the Seebeck coefficient  are 
remarkably different in each sample, ranging from -500 $\mu$V/K to -45 mV/K.\cite{FeSb2,FeSb5,FeSb9,FeSb10,FeSb11}
A systematic study of the impurity effects has revealed that 
the Seebeck coefficient is simply related to the carrier concentration not to the electron correlation.\cite{FeSb6}
Moreover, the low-temperature magnetotransport behaviors are well
understood as that of an extrinsic semiconductor with ppm-level impurity.\cite{FeSb13}
To account for the observed large Seebeck coefficient, a phonon-drag mechanism has been suggested,\cite{FeSb12}
as generally seen in conventional semiconductors.

In heavy-fermion compounds including Kondo insulators, localized moments become itinerant through the Kondo effect, where a typical temperature called the Kondo temperature is proportional to $\exp(-1/J)$ with the exchange energy $J$.
In Kondo insulators, the gap size also scales with the Kondo temperature.
An important feature is that $J$ is roughly proportional to $V^{2}$, where $V$ is the transfer integral between the localized and conduction electrons.\cite{CeBiPt, Normand}
Consequently, the gap of the Kondo insulators is extremely sensitive to lattice compression,\cite{Doniach} whereas a conventional semiconductor gap is proportional to $V$, giving a moderate pressure dependence. 
For example, in the Kondo insulator Ce$_{3}$Bi$_{4}$Pt$_{3}$, 
the application of physical pressure results in an increase in the gap size according to the expectation of hybridization gap model.\cite{CeBiPt}
In SmB$_6$ and YbB$_{12}$, 
the gap magnitude is also easily modified through the hybridization change by pressure but the opposite effect has been observed 
as the gap is suppressed with pressure,\cite{Bille83,Iga93}
which is possibly attributed to the ion-size difference between various $4f$ configurations.\cite{Xu96}

Here, to clarify the origin of the gap formation mechanism in FeSb$_2$,
we show the hydrostatic pressure effect on two energy gaps through the electrical resistivity and magnetotransport measurements.
In this compound, chemical pressure due to element substitution cannot examine the effect of the hybridization on the gap formation precisely, 
since the magnetic and transport properties are sensitively varied by a ppm-level impurity concentration.\cite{FeSb6} 
In contrast, physical pressure can reduce the unit cell volume, leaving impurities/defects intact.
We evaluate the larger and the smaller gaps from the resistivity and the magnetotransport properties, respectively, in various pressures.
Both the energy gaps are found to increase with pressure, but the increase rate differs each other:
the pressure variation of the smaller gap $\Delta_2$ is weak, compared with that of the larger gap $\Delta_1$.
This pressure-insensitive energy gap is compatible with a simple band picture that $\Delta_{2}$ is a gap formed between an impurity level and a conduction band.

\section{Experiment}
High-quality single crystals of FeSb$_{2}$ were grown by a self-flux method using metal powders of 99.999 $\%$ (5N) pure Fe and 99.9999 $\%$ (6N) pure Sb, and detailed information was given in Ref.\:\onlinecite{FeSb6}.
The magnetic impurity was evaluated to be less than 0.01 \%.
As shown in Fig. 1(a), the Seebeck coefficient changes the sign from positive to negative with decreasing temperature around 40 K, which is a sign for high-quality.\cite{FeSb2, FeSb6, FeSb13}
It reaches $-1400$ $\mu$V/K, which also evidences the quality.\cite{FeSb6, FeSb13, Qing}
An as-grown crystal of $3\times 3\times 3$ mm$^{3}$ was cut into a rectangular piece of $0.5\times 0.7\times 1$ mm$^{3}$. 
The resistivity was measured from 300 down to 3 K.
The magnetic field dependence of electrical and Hall resistivities was measured up to 70 kOe in the temperature range from 30 down to 5 K.
These transport properties were measured with a conventional four-probe dc method using Physical Property Measurement System (Quantum Design, Inc.) using a piston cylinder clamp cell made of BeCu (ElectroLAB company).
Daphne oil 7373 was used as a pressure-transmitting medium and the applied pressure was evaluated from the superconducting transition temperature of Pb placed near the sample. 
\section{Results and Discussion}
Figure 1 shows the temperature dependence of the resistivity $\rho$ below 300 K at 0, 0.9 and 1.8 GPa.
The resistivity at 0 GPa exhibits insulating behavior with two upturns separated by a plateau near 20 K, indicating the existence of the two energy gaps in FeSb$_{2}$.\cite{FeSb2,FeSb5,FeSb9,FeSb10} 
The plateau temperature increases from 20 to 30 K with pressure, and the magnitude of $\rho$ also increases with pressure above 50 K.
We evaluate the large energy gap $\Delta _{1}$ and small energy gap $\Delta _{2}$ at each pressure with a thermal activation function of $\rho=\rho _0\exp(\Delta /2k_{B}T)$.
As shown in Fig. 1(b), slopes of the dotted lines above 50 K and below 10 K in the Arrhenius plot give a values of $\Delta_{1}$ and $\Delta _{2}$, respectively.
Clearly, the slope above 50 K increases with pressure, meaning that the energy gap $\Delta _{1}$ is enhanced from 30 to 40 meV.
The enhancement in $\Delta_{1}$ is consistent with that for a polycrystalline sample reported by Mani $et$ $al$.\cite{FeSb14}
On the other hand, the slope change below 10 K is smaller than that above 50 K, suggesting a weak pressure dependence of $\Delta _{2}$ (5 meV at 0 GPa and 7 meV at 1.8 GPa), which is not visible in the polycrystalline sample.
However, to evaluate the pressure dependence of $\Delta _{2}$, there are two important problems. 
One is that the carrier mobility strongly increases with decreasing temperature below 30 K, reported in Ref. 11, indicating that the energy gap cannot be precisely determined with the resistivity.
The other problem is that we should evaluate the energy gap at the low temperature range, but the resistivity saturates below 4 K due to the impurity band conduction.\cite{FeSb2,FeSb6,FeSb13}
As a result, we can only obtain the gap size with the fitting line at the narrow temperature range as shown in Fig. 1(b), leading an inappropriate estimation of the gap size.
Thus, we precisely evaluate the pressure dependence of the energy gap $\Delta_{2}$ using the temperature dependence of carrier concentration under pressure  with the magnetotransport properties as shown below.

\begin{figure}[t]
\begin{center}
\includegraphics[width=8cm]{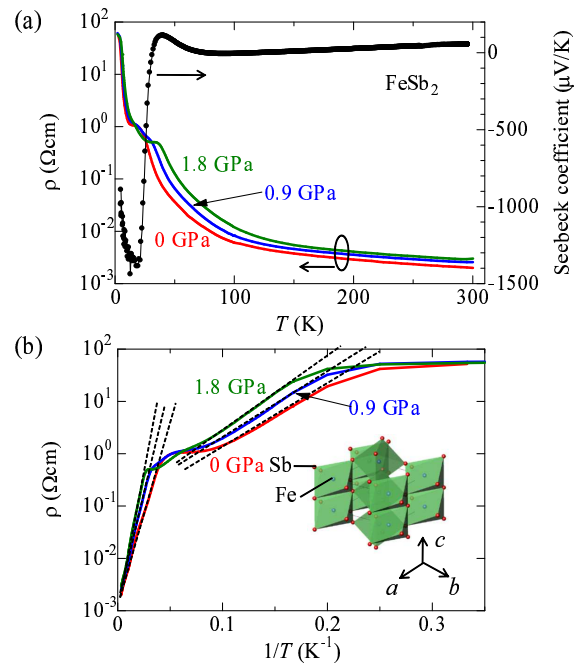}
\caption{(Color online) (a) The temperature dependence of the resistivity $\rho$ at various pressure (left axis) and the temperature dependence of the Seebeck coefficient (right axis). (b) The resistivity as a function of 1/$T$, where energy gaps are evaluated by an activation function $\rho\propto\exp(\Delta /2k_{B}T)$, plotted as dotted lines. In the inset of (b), the crystal structure of FeSb$_{2}$ is schematically shown.}
\end{center}
\end{figure} 

In Figs. 2(a)-(f), we show the magnetic field dependence of the Hall resistivity $\rho_{xy}$ and the transverse magnetoresistance $\Delta  \rho/\rho_{0} \equiv [\rho_{xx}(H) -\rho_{xx}(0)]/\rho_{xx}(0)$, measured at various temperatures from 5 to 30 K at 0, 0.9 and 1.8 GPa.
At ambient pressure, the field dependence of $\rho_{xy}$ and $\Delta \rho/\rho_{0}$ is almost the same as that of the previously report.\cite{FeSb13}
$\rho_{xy}$ and $\Delta \rho/\rho_{0}$ at high pressure are qualitatively similar to those at ambient pressure, while quantitative values are changed with pressure.

\begin{figure}[t]
\begin{center}
\includegraphics[width=8cm]{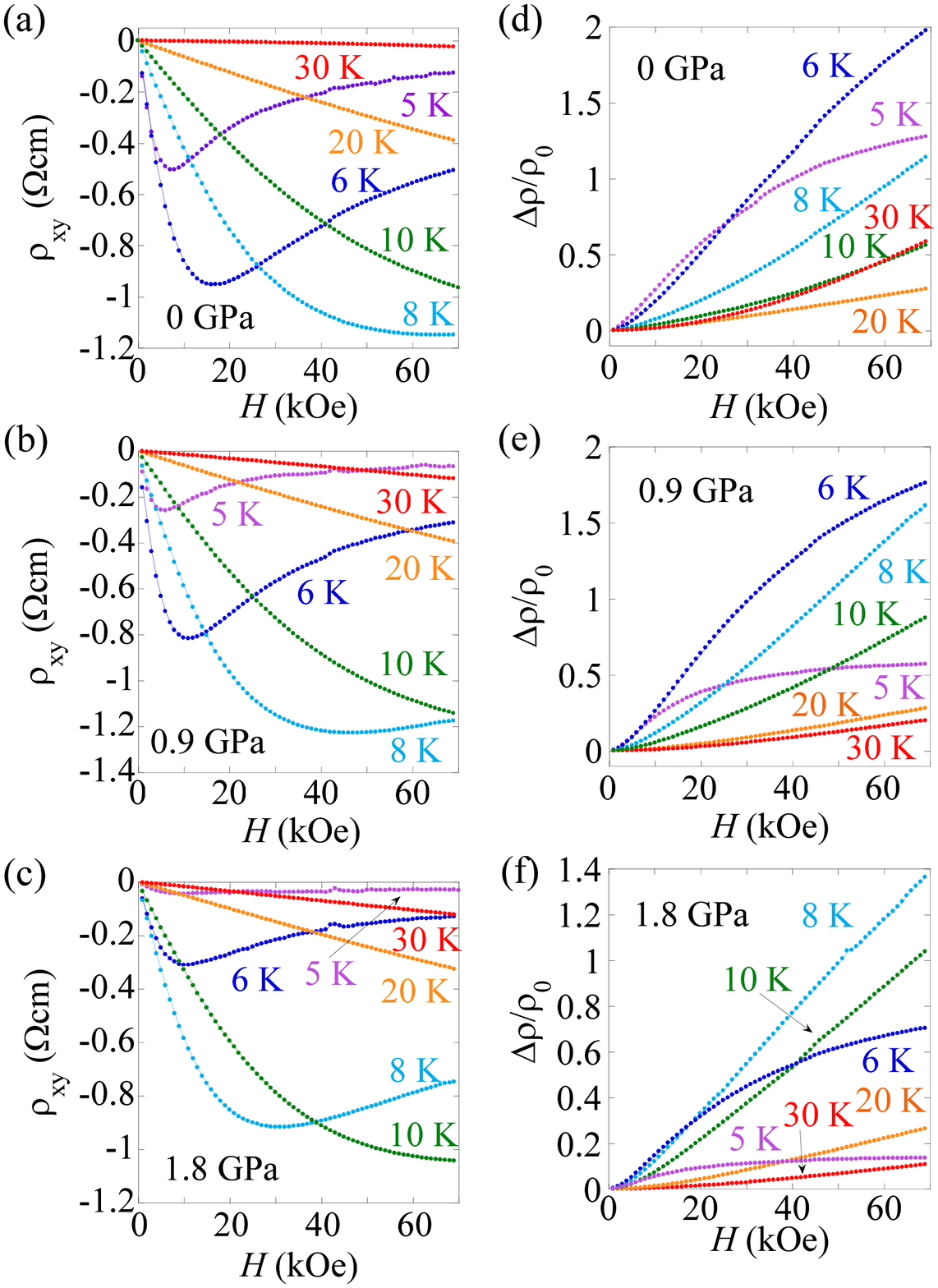}
\caption{(Color online) (a)-(c) The Hall resistivity $\rho_{xy}$, and (d)-(f) the transverse magnetoresistance $\Delta  \rho/\rho$ as a function of magnetic field measured at various pressure.}
\end{center}
\end{figure}

According to Ref. 20, let us evaluate the carrier concentration and the mobility. 
The unique field dependence of $\sigma _{xy}=-\rho_{xy}/(\rho_{xx}^{2}+\rho_{xy}^{2})$, is obtained at various pressure, as shown in Figs. 3(a)-(f). 
We fit these data on the basis of a two-carrier model, where we assume that one carrier is of high mobility and the other is of low mobility ($\mu_{\rm low} H\ll 1$), described as
\begin{equation}
\sigma_{xy} (H)=ne\mu ^{2}H\left[\frac{1}{1+(\mu H)^{2}}+C\right].
\label{eq:sxy}
\end{equation}
Here, $n$, $\mu$, and $C$ are the carrier concentration of the high-mobility component, the carrier mobility of the high-mobility component, and the low-mobility component for $\sigma_{xy}$, respectively.
The best agreements of $\sigma_{xy}$ with Eq. (\ref{eq:sxy}) are shown by the solid curves in Fig. 3.
Note that the fitting curves well explain the experimental data at all the magnetic field range.
Accordingly we do not need to consider magnetic field dependence in fitting parameters.
The mobility estimated from $C$ is about 15 \% of the high-mobility component at all temperatures, and therefore, $\mu_{\rm low} H \ll 1$ holds, and our model is valid for all the magnetic fields.

\begin{figure}[t]
\begin{center}
\includegraphics[width=8.5cm]{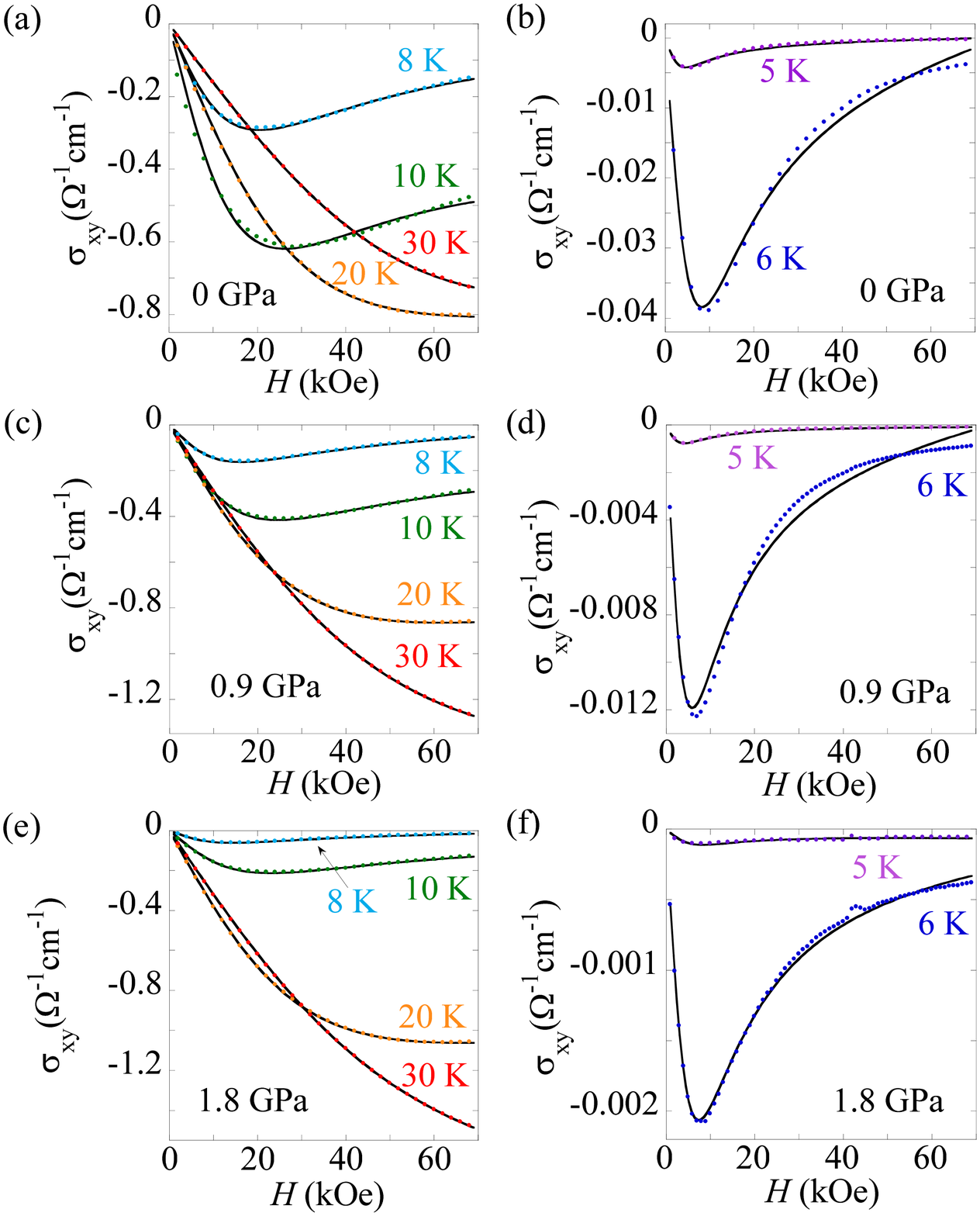}
\caption{(Color online) (a)-(c) The magnetic field dependence of the conductivity tensor $\sigma_{xy}$ at various pressure from 5 to 30 K. Solid curves are the calculation using Eqs. (\ref{eq:sxy}).}
\end{center}
\end{figure}

Using the above analysis, we plot the temperature dependence of the carrier mobility and the carrier concentration in Figs. 4(a) and 4(b).
The mobilities increase with decreasing temperature from $\sim1000$ cm$^{2}$/Vs at 30 K to $\sim15000$ cm$^{2}$/Vs at 5 K and have almost the same values at all pressure range.
The carrier concentration shows semiconducting-like behavior, from which we evaluate the small energy gap $\Delta _{2}$ using an activation function of $n=n_{0}\exp(-\Delta _{2}/2k_{B}T)$, described as the dotted lines in Fig. 4(b).
The extrapolated lines using the data below 10 K deviate from the data above 20 K, because $\Delta_{2}/2k_{B}$ is of the order of 10 K.
Above 10 K, the thermal fluctuation becomes comparable to the gap energy, and hence $n$ does not follow the activation function.
In Fig. 4(c), we plot the pressure dependence of $\Delta _{1}$ and $\Delta _{2}$, showing that energy gaps $\Delta _{1}$ and $\Delta _{2}$ increase by 10 meV  and 3.5 $\pm$ 1.5 meV in going from 0 to 1.8 GPa, respectively.
Here, the error bars of $\Delta_{2}$ in Fig. 4(c) come from a difference of a fitting temperature range.
Maximum values of the energy gap at each pressure are evaluated with the slope between 8 and 5 K, while minimum values are evaluated with the slope between 10 and 5 K.
In this method, the maximum and minimum values at 0 GPa, for example, are 13.4 and 10 meV, respectively, and then the energy gap can be plotted as 11.7 $\pm$ 1.7 meV. 
The resistivity of high mobility component $\rho_{\rm high}$ ($= 1/ne\mu$) is presented in Fig. 4(d), pressure dependence of which is almost consistent with the resistivity data of Fig. 1(b), indicating that the transport properties below 30 K is dominated by high mobility component. However, the value of $\rho_{\rm high}$ at 5 K is a little higher than that of Fig. 1(b), which suggests that low mobility components such as impurity band carriers affect the transport properties below 5 K (see the later discussion).\cite{FeSb13}

\begin{figure}[t]
\begin{center}
\includegraphics[width=8cm]{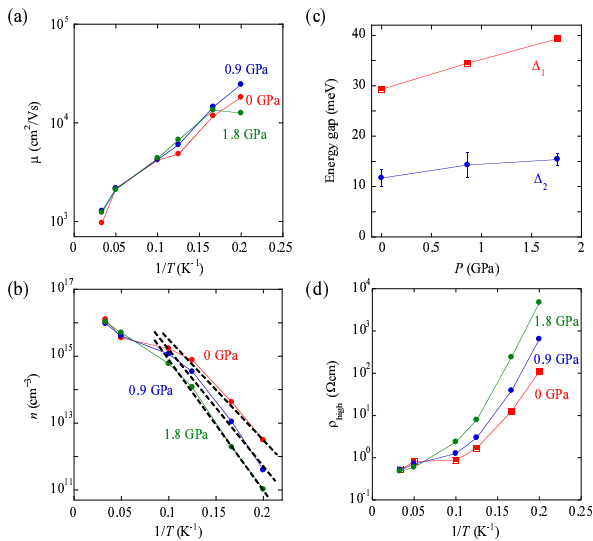}
\caption{(Color online) 1/$T$ dependence of (a) the carrier mobility $\mu$ and (b) the carrier concentration $n$, obtained from the fitting results for $\sigma_{xy}$. The dotted lines are the fitting results to evaluate the low temperature energy gap by the thermal activation function $n\propto\exp(-\Delta_{2}  /2k_{B}T)$.  (c) The pressure dependence of energy gaps of $\Delta_{1}$(red square) and $\Delta_{2}$(blue circle). (d) The resistivity of high-mobility component $\rho_{\rm high} (=1/ne\mu $).}
\end{center}
\end{figure}

Here, we attempt to understand an origin of the pressure dependence of $\Delta_{1}$.
The gap enhancement is expected from a band calculation by Wu $et$ $al$; the gap of 0.2 eV at ambient pressure increases by 0.25 eV at 5 GPa.\cite{FeSb15}
Conventional semiconductors show the gap enhancements of order of 10 meV/GPa, since the periodic potential of electron in these material can vary by the lattice compression/expansion.\cite{p-semi, p-semi2}
These relative changes of the energy gaps are almost the same as that of our sample, suggesting that the enhancement of $\Delta_{1}$ can be explained by a simple band model.
On the other hand, this enhancement can be accounted for a hybridization between the narrow $d$-band and the broad conduction band due to the Kondo mechanism.
In the Kondo system, the gap size is modified by changing the strength of hybridization $V$.
Applying pressure, $V$ tends to increase by approaching the narrow $d/f$ band to the conduction band, as a result the energy gap $\Delta$ increase, since $V$ dependence of the gap size can be described as $\Delta\propto \mid V\mid ^2 $.\cite{FeSb14,KondoP1,KondoP2}
The relative change of gap size enhancement of $\Delta_{1}$ is almost the same as that of FeSi, which is considered as $d$-electron Kondo insulator.\cite{FeSi4}
The relation between the strongly correlated effects and the gap formation mechanism of FeSb$_{2}$ was suggested by an optical measurement, which observed an unusual behavior that an opening of the gap is associated with a large transfer of the spectral weight over a energy range of $\sim$ 1 eV compared with the value of the energy gap.\cite{FeSb4}
From the above results, the pressure effect cannot directly clarify that $\Delta_{1}$ is formed by the simple band model or an electron correlation effect such as the Kondo mechanism.

An origin of the small gap $\Delta_{2}$ is important because some groups think that the huge Seebeck coefficient is observed due to the small energy gap of $\Delta_{2}$ opened by strong correlation effects.\cite{FeSb2,FeSb4,FeSb11}
The energy gap increases from 12 meV at 0 GPa to 15 meV at 1.8 GPa, and the increment in $\Delta_{2}$ is less than half that in $\Delta_{1}$.
Here, according to several band calculation studies,\cite{FeSb11,FeSb12} the bands near the Fermi level are mainly formed by $t_{2g}$ orbitals of Fe $d$ electrons, which are split into two occupied lower-lying levels with $d_{yz}$ and $d_{zx}$ symmetries and one unoccupied higher-lying level with $d_{xy}$ symmetry in a simple crystal-field picture.
If $\Delta_{2}$ is formed by such a crystal field splitting of $d$ orbitals, $\Delta_{1}$ may possess another orbital character such as $s$- or $p$-orbitals.
In this case, the pressure dependence of $\Delta_{2}$ will be larger than that of $\Delta_{1}$, because such a crystal-field splitting sensitively increases with pressure.
We also discuss the Kondo-insulator scenario, in which the gap will be quite sensitive to the high pressure compared with the gap of the single particle band structure.
Actually, a typical $f$-electron Kondo insulator Ce$_{3}$Bi$_{4}$Pt$_{3}$, which has a Kondo gap of 50 K at ambient pressure,\cite{CeBiPt} shows more sensitive pressure dependence of the energy gap.
This Kondo gap becomes 100 K at 10 kbar.
The small pressure effect of $\Delta_{2}$ seems to indicate that $\Delta_{2}$ does not come from the Kondo effect.
Rather the present result is consistent with the previous study, where we suggested that $\Delta_{2}$ is a gap between the impurity band and the conduction band.\cite{FeSb13}
The magnitude of the carrier concentration $n=n_{0}\exp(-\Delta_{2}/2k_{B}T)$ also indicates that the gap $\Delta_{2}$ is unlikely to come from the Kondo gap.
As shown in Fig. 4(b), $n_{0}$ falls on $10^{18}-10^{19}$ cm$^{-3}$, which corresponds to 0.01 carrier per unit cell.
The value of $n_{0}$ roughly indicates a saturated carrier concentration at high temperature, and accordingly the small $n_{0}$ implies that very few carriers are available.
In Kondo insulators, $n_{0}$ corresponds to the density of states of the conduction band, being $10^{2}-10^{3}$ times larger.\cite{Hiraoka}

Considering the above analysis, we depict a schematic band picture of FeSb$_{2}$ in Fig. 5, showing that $\Delta_{1}$ is a gap between the conduction band and the valence band, and the impurity band exists in the band gap, and $\Delta _{2}$ is a gap from the impurity band to the conduction band.
In this case, $\Delta _{2}$ increases with increasing high-temperature gap.
This model can qualitatively explain the smaller pressure dependence of $\Delta _{2}$, inferring that $\Delta_{2}$ is not directly related to the Kondo gap in this system.
This model, however, cannot reproduce the value of the huge Seebeck coefficient at low temperature with the non-degenerate band model equation described as $|S|\propto k_{B}/e(\Delta_{2} /2k_{B}T)$; $S$ becomes 1 mV/K at 10 K at most.
For this reason, an additional term such as the phonon drag may be important for the huge Seebeck coefficient.  

\begin{figure}[t]
\begin{center}
\includegraphics[width=8cm]{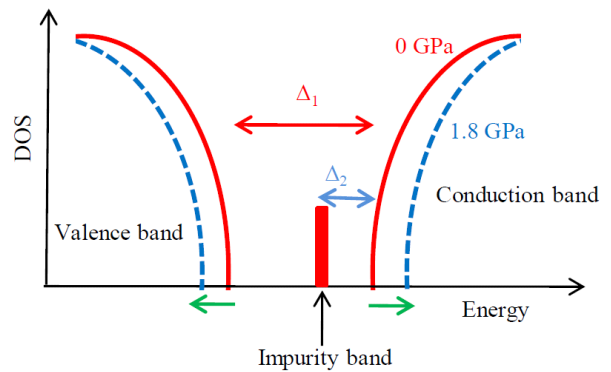}
\caption{(Color online) The schematic band picture of FeSb$_{2}$. The impurity band exists between the conduction and valence band. $\Delta_{1}$ is enhanced because the conduction and valence band energy shift due to the change of the hybridization strength between the narrow $d$ band and broad $p$ band. The low energy gap also increases with the energy shift of the conduction band.}
\label{fig:uxxuxy}
\end{center}
\end{figure}        
 
\section{conclusion}
We have measured the temperature and magnetic field dependence of the electrical and Hall resistivity of high-quality FeSb$_{2}$ single crystals at various pressures, and analyzed the pressure dependence of the high and low energy gaps.
The large energy gap $\Delta _{1}$ increases from 30 to 40 meV in going from 0 to 1.8 GPa, which implies that hybridization between the narrow $d$ bands and the broad $p$ bands is important for $\Delta _{1}$.
On the other hand, a weak pressure dependence of the small gap $\Delta_{2}$ is observed.
This weak pressure dependence suggest that the origin of the gap is not associated with strongly correlated effect such as Kondo effect, because an typical 4$f$-Kondo insulators Ce$_{3}$Bi$_{4}$Pt$_{3}$ shows a sensitive pressure effect that the relative enhancement of the gap is 5 times larger than that of our sample.
We propose a band picture that $\Delta_{2}$ is a gap between an impurity level and a conduction band, which naturally explains the transport properties and the pressure dependence.
We also propose that the Kondo effect is unlikely to be the origin for the huge Seebeck coefficient.

\section*{Acknowledgements}
The authors would like to thank M. Sato for the initial motivation of this work. 
This work was partially supported by Program for Leading Graduate Schools, Japan Society for Promotion of Science and by Advanced Low Carbon Research and Development Program, Japan Science and Technology Agency.

\end{document}